\documentclass[aps,pre,showpacs,twocolumn,groupedaddress]{revtex4-1}

\usepackage{bm}
\usepackage{psfrag}
\usepackage{graphicx}
\usepackage{amsfonts}
\usepackage{amsmath}
\usepackage{amssymb}
\usepackage{dcolumn}

\begin{document}


\title{Randomly Evolving Idiotypic Networks:\\Modular Mean Field Theory}


\author{Holger Schmidtchen}
\author{Ulrich Behn}
\email[]{ulrich.behn@itp.uni-leipzig.de}
\affiliation{Institut f\"ur Theoretische Physik, Universit\"at
  Leipzig, POB~100~920, D-04009~Leipzig, Germany}
\affiliation{International Max Planck Research School Mathematics in the
  Sciences,\\Inselstra\ss e 22, D-04103 Leipzig, Germany}


\date{\today}

\begin{abstract}

We develop a modular mean field theory for a minimalistic model of the idiotypic network. The model comprises the random influx of new idiotypes and a deterministic selection. It describes the evolution of the idiotypic network towards complex modular architectures, the building principles of which are known. The nodes of the network can be classified into groups of nodes, the modules, which share statistical properties. Each node experiences only the mean influence of the groups to which it is linked. Given the size of the groups and linking between them the statistical properties such as mean occupation, mean life time, and mean number of occupied neighbors are calculated for a variety of patterns and compared with simulations. For a pattern which consists of pairs of occupied nodes correlations are taken into account.

\end{abstract}




\maketitle

\section{Introduction\label{sec:introduction}}


B-Lymphocytes express on their surface Y-shaped receptors, called antibodies, with highly specific binding sites. All antibodies of a given B-cell are of the same type, the idiotype. If the antibodies are cross-linked by complementary structures, situated e.g. on an antigen, the B-cell is stimulated to proliferate and, after a few cell cycles, differentiate into plasma cells and memory cells. Plasma cells secrete large amounts of antibodies, which attach to the antigen and mark it for further processing. Useful clones survive, while others, lacking stimulation, die \cite{Burnet59}.

Complementary structures can be found also on B-lymphocytes. B-cells of complementary idiotype may stimulate each other, thus the B-lymphocyte system forms a functional network, the idiotypic network \cite{Jerne74}. A history and thorough discussion of
immunological paradigms can be found in \cite{Carneiro97}, cf. also \cite{Coutinho03}. For reviews on idiotypic networks with emphasis on modeling see \cite{Behn07}, and with focus on new immunological and clinical findings see \cite{Behn11}.

The idiotypic network is an attractive concept for system biologists, but due to their complexity also a challenge for theoretical physicists. The size of the potential idiotypic repertoire of humans is estimated to exceed $10^{10}$ \cite{BM88}, the expressed repertoire is of order $10^8$ \cite{PW97}. Interactions between B-cells of complementary idiotype are genuinely nonlinear. Thus, modeling idiotypic networks is an inviting playground for statistical physics, nonlinear dynamics, and complex systems.  Networks, especially random and randomly evolving networks, with applications in a plethora of different, multidisciplinary fields \cite{Strogatz01,DM03,Caldarelli07,GS09,Newman10} experience great interest in the community of statistical physicists. Computer scientists try to mimick the immune system to fight against foreign invaders \cite{FB07}.


A minimalistic model of the idiotypic network was proposed in \cite{BB03} where the nodes represent B-lymphocytes and antibodies of a given idiotype. The idiotype is characterized by a bitstring. Populations with complementary idiotypes, allowing for a few mismatches, can interact. In the model, an idiotype population may be present or absent. For survival it needs stimulation by sufficiently many complementary idiotypes, but becomes extinct if too many complementary idiotypes are present. This reflects the log-bell shaped dose-response curve characteristic for B-lymphocytes \cite{PW97}. The dynamics is driven by the influx of new idiotypes generated by mutations in the bone marrow.

The potential idiotypic network consists of all idiotypes an organism is able to generate. Each idiotype $v$ is labelled by a bitstring of length $d$: $\bm{b}_d \bm{b}_{d-1} \cdots \bm{b}_1$, which is the binary address of the node in the network.
Two nodes $v$ and $w$ are linked if their bitstrings are complementary. We allow for $m$ mismatches, i.e. their Hamming distance must obey $d_H(v,w)\geqslant d\!-\!m$. These nodes and links build an undirected graph $G_d^{(m)}$, the base graph. Each node has the same number of neighbors, $\kappa=\sum_{k=0}^m {d \choose k}$. The expressed idiotypic network is only a part of the potential network.

New idiotypes generated in the bone marrow are introduced by occupying empty nodes randomly. Occupied nodes are selected to survive if they receive sufficient stimulus, i.e. if the number of occupied neighbors is within an allowed window. To be specific, the rules for (parallel) update are
\begin{enumerate}
\vspace{-1mm}
\item[(i)] Occupy empty nodes with probability $p$\\[-6.5mm]
\item[(ii)] Count the number of occupied neighbors $n(\partial v)$ of node $v$. If $n(\partial v)$ is outside the window $[t_L, t_U]\,$, set the node $v$ empty\\[-6.5mm]
\item[(iii)] Iterate.\\[-5.5mm]
\end{enumerate}

The model has a minimal number of parameters, namely the length of the bitstring, the allowed number of mismatches, upper and lower thresholds of the window, and the influx probability of new idiotypes.


We can consider our model system as a descendant of Conway's game of life \cite{Gardner70}. There, on an infinite regular 2d lattice the sites can take value 0 or 1. The system is updated in parallel in discrete time.
If the number of living neighbors lies between a lower and an upper threshold, a site becomes populated or survives in the next step. The dynamics is entirely determined by the initial configuration. There is continuous interest in the highly complex properties of game of life, for a recent status report see \cite{Adamatzky10}.

Schulman and Seiden \cite{SS78} investigated a probabilistic version of the game of life, where the update rule is modified in two ways. Sites for which a window rule is fulfilled will be occupied or will survive with given probabilities. Furthermore, a site can be occupied or survives in a stochastic way, parametrized by a temperature $T$ such that for $T=0$ the modified window rule applies but plays no role for $T\to \infty$. Starting from random initial conditions, simulations show a sharp transition of the global mean occupation when $T$ is increased. The high temperature phase is well described by a mean field theory which fails however to reproduce low temperature results. Excluding sites without occupied neighbors improves the agreement of theory and simulation for low temperatures. A second order mean field theory was proposed in \cite{BRR91}.

Gutowitz et al. \cite{GVK87} developed a local structure theory for cellular automata on regular lattices, which considers the evolution of joint probabilities of sets of neighboring cells (block configurations) and thus include correlations. Application to the problem of Schulman and Seiden yields results which agree with their simulations \cite{GV87}.

Bidaux et al. \cite{BBC89} considered a binary probabilistic cellular automaton with a totalistic update rule involving 8 neighbors on regular lattices in $d=1, \dots, 4$. If the window rule is fulfilled, it applies with probability $p$. Simulations with random initial conditions show a transition of the global mean occupation at a critical value $p_c$, which is continuous for $d=1$ and first order for $d>1$. A mean field theory describes qualitatively a first order transition. An overview of mean field theories of cellular automata including the probabilistic game of life can be found in \cite{Ilachinsky01}

In the context of network models there are numerous mean field approaches aimed to describe degree distribution, clustering coefficient, average shortest distance between two nodes, mean populations, e.g. in susceptible-infectious-recovered models of epidemic spread, and other characteristics. Many references can be found in the recent monograph by Newman \cite{Newman10}, cf. also \cite{dSS05,BLMCH06,DM06,Caldarelli07}.

Gleeson and Cahalane \cite{GC07} investigate cascades of activation in a model with a threshold dynamics on a Poisson random graph with average degree $z$. Starting with few active nodes, neighbors become permanently active if the fraction of active neighbors exceeds a threshold, drawn for each node from a given distribution. The fraction of activated nodes in the $n$th update step is determined recursively. The final fraction of active nodes $\rho$ is related to the fixed points of this recursion. For Gaussian distributed thresholds with mean $R$ this fraction as a function of $z$ may change in a continuous or discontinuous way depending on $R$.

In this paper we deal with a network of complex architecture which emerges as a result of a random evolution. The architecture is build of groups of nodes, the modules, which share statistical properties.  To describe this architecture, a global mean field theory is obviously not appropriate. Depending on the parameters, different architectures can occur. In a previous paper \cite{STB} we have reached a detailed understanding of the building principles of the architectures, which allows to calculate the number and size of the groups and their linking. Here we develop a modular mean field theory to calculate the statistical properties for given architectures.


In the next section we shortly sketch the building principles \cite{STB} and collect the results needed to develop the theory. In Sec.~\ref{sec:meanfield} we determine the evolution equation for the mean population of the groups for a given architecture, which properly takes into account the update rule of the model, random influx and the window rule. The fixed point of this evolution equation gives the stationary mean populations and allows to compute also the mean life time and the mean occupation of the neighborhood. In Sec.~\ref{sec:specialcases} we consider two groups of nodes for which the mean field theory considerably simplifies: Singletons, which are essentially isolated nodes, and groups of self-coupled nodes, core groups, in static patterns. In Sec.~\ref{sec:correlation} we extend the mean field theory by including correlations, which is necessary to describe 2-cluster patterns. The results of the mean field theory for the general case are compared with simulations for a range of the influx parameter $p$ in Sec.~\ref{sec:evaluation}. Some probabilistic aspects of the stability of patterns are discussed in the appendix. 

\section{Architecture of Patterns\label{sec:architecture}}

Simulations of the model for one and two allowed mismatches revealed that the system evolves for typical parameters towards a complex functional architecture \cite{BB03}. Groups of  nodes were identified which share statistical properties such as the mean occupation, the mean life time and the mean number of occupied neighbors. Also the size of the groups and the linking between them have been determined. With increasing influx of new idiotypes transitions between architectures of different complexity are observed. For small influx static patterns are found. For an intermediate range of the influx a stationary dynamic architecture is observed which is the most interesting one. It includes a densely connected core, a periphery, and isolated nodes (singletons), resembling the notion of central and peripheral part of the biological network \cite{Coutinho89,VC91}. For larger influx the architecture becomes more irregular.

In \cite{STB,SB06,*SB08} an analytic description of the general building principles of these architectures was proposed. It allows to calculate number and size of groups and their linking for a given architecture. Ideal static patterns, i.e. patterns without defects which persist without influx, are completely characterized.

For a given architecture, i.e. a pattern, the nodes can be classified according to the values of bits in determinant positions common to all nodes. Different patterns are characterized by the number $d_M$ of such positions. The entries in these determinant positions decide to which group a node belongs. The pattern can be built by regular arrangement of elementary building blocks, which are hypercubes of dimension $d_M$. We call these building blocks pattern modules. The concept is explained in detail and many examples are given in \cite{STB}.

For a given pattern, i.e. a given $d_M$, the number of groups and their sizes can be calculated combinatorially
\begin{equation}
  |S_g| = 2^{d-d_M} {d_M \choose g\!-\!1} \,,\ g = 1, \dots, d_M\!+\!1\,,
  \label{eq:groupsize}
\end{equation}
and the number of links $L_{gl}$ of a given node $v_g \in S_g$ to nodes in $S_l$ is
\begin{align}
  L_{gl} = & \sum_{i,j} {d\!-\!d_M \choose j} {g-1 \choose
    i\!+\!\max(0,-\Delta_{gl})}  \nonumber \\
  & \times {d_M-g+1 \choose i+\max(0,\Delta_{gl})} \openone
  (j+2i+|\Delta_{gl}| \leqslant m) \label{eq:linkmatrix}
\end{align}
where $\Delta_{gl} = d_M\!\!-\!g\!-\!l\!+\!2$. 

For example, on a base graph $G_{12}^{(2)}$ with $[t_L, t_U] = [1, 10]$ we observe an architecture of three groups. It has two determinant bits and is described by a pattern module of dimension $d_M=2$. Persisting occupied nodes in the first group form 2-clusters, potential hubs, the second group, occasionally link together several 2-clusters, and the nodes in the third group are stable holes.  The links are given by the corresponding matrix $\mathbb{L}=(L_{gl})$.

\begin{figure}[t]
\centering
\includegraphics[width=\columnwidth]{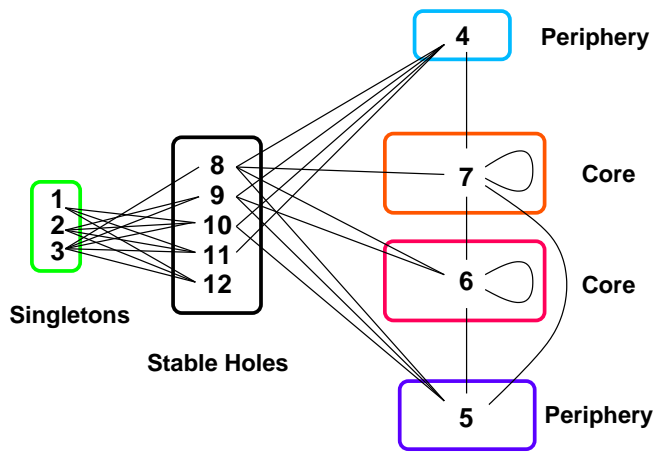}
\caption{(Color online) A visualization of the 12-group architecture,
  $d_M=11$, observed in simulations for $p \approx 0.025 \dots 0.045$ on a base graph
  $G_{12}^{(2)}$ with $[t_L, t_U]=[1,10]$ in agreement with Eqs.~(\ref{eq:groupsize}) and (\ref{eq:linkmatrix}). The size of the boxes
  corresponds to the group size.  The lines show possible links
  between nodes of the groups.}
 \label{fig:scheme12g}
\end{figure}

On the same base graph, for intermediate influx $p$, we also find a dynamical 12-group architecture. Its architecture, visualized in Fig.~\ref{fig:scheme12g}, is based on a pattern module of dimension $d_M=11$. The corresponding link matrix is given in Table~\ref{tab:linkmat12g}. We find two large core groups with links to almost all other groups, two peripheral groups connected to the core, groups of stable holes, which separate the singletons from the central network. 

In both cases the architecture is stationary, simulation and analytic predictions agree perfectly.

\begin{table}
\centering
\caption{Linkmatrix and qualitative classification of groups for the
  12-group architecture of Fig.~\ref{fig:scheme12g}. Missing entries are zero.\\ \label{tab:linkmat12g}} 
\renewcommand{\arraystretch}{1.3}
\begin{tabular}{|lc|*{3}{c}|*{2}{c}|*{2}{c}|*{5}{c}|}
  \hline
  & & \multicolumn{3}{c|}{singletons} & \multicolumn{2}{c|}{periph} &
  \multicolumn{2}{c|}{core} & \multicolumn{5}{c|}{stable holes} \\
  & & $S_{1}$ & $S_{2}$ & $S_{3}$ & $S_{4}$ & $S_{5}$ & $S_{6}$ & $S_{7}$
  & $S_{8}$ & $S_{9}$ & $S_{10}$ & $S_{11}$ & $S_{12}$ \\ 
  \hline
  & $v_{1}$ & &  &  &  &  &  &  &  &  &  55 & 22& 2 \\
  singletons & $v_{2}$ & &  &  &  &  &  &  &  &  45& 20 & 12& 2 \\ 
  & $v_{3}$ & &  &  &  &  &  &  &  36& 18& 20 & 4 & 1 \\ 
  \hline
  & $v_{4}$ & &  &  &  &  &  &  28& 16& 26& 6& 3 &  \\
  \raisebox{1.65ex}[-1.65ex]{periphery} & $v_{5}$ & &  &  &  &  &  21&
  14& 30& 8&  6&  &  \\  
  \hline 
  & $v_{6}$ & &  &  &  &  15& 12& 32& 10& 10& &  &  \\ 
  \raisebox{1.65ex}[-1.65ex]{core} & $v_{7}$ & &  &  &  10& 10& 32& 12& 15& &  &  &  \\ 
  \hline
  & $v_{8}$ & &  &  6&  8&  30& 14& 21& &  &  &  &  \\ 
  & $v_{9}$ & &  3&  6&  26& 16& 28& &  &  &  &  &  \\ 
  st. holes & $v_{10}$ & 1&  4&  20& 18& 36& &  &  &  &  &  &  \\ 
  & $v_{11}$ & 2&  12& 20& 45& &  &  &  &  &  &  &  \\ 
  & $v_{12}$ & 2&  22& 55& &  &  &  &  &  &  &  &  \\
  \hline
\end{tabular}
\end{table}

The occupation of nodes is in general not permanent but fluctuates due to the influx while the architecture persists. For a given pattern the mean occupation of the groups varies systematically with the influx $p$. The mean occupation and other statistical quantities can be computed in a modular mean field theory. We consider the nodes as situated in a mean field exerted by the neighboring groups characteristic for a given architecture. The required link matrix $\mathbb{L}$ in our model can be calculated or obtained in simulations. Other modular architectures could be treated, too.

\section{Mean Field Approach\label{sec:meanfield}}

\newcommand{\Prob}{\text{Prob}}
\newcommand{\tildeC}{\widetilde{C}}
\newcommand{\tildeP}{\widetilde{P}}
\newcommand{\tildekl}{\widetilde{k}_l}
\newcommand{\Lgl}{L_{gl}}
\newcommand{\tilden}{\widetilde{n}}
\newcommand{\tildev}{\tilde{v}}
\newcommand{\tildew}{\tilde{w}}
\newcommand{\PW}[1]{P^{\mathbb{W}}_{#1}}
\newcommand{\dvg}{\partial v_g}
\newcommand{\dvtg}{\partial \tilde{v}_g}
\newcommand{\dvpg}{\partial v'_g}

\subsection{Mean Occupation\label{sec:meanoccupation}}

In a mean field approach we assume that all nodes are independently
occupied with a probability $\Prob(n(v_g)\!=\!1)\,$, $v_g \in
S_g$, characteristic for the group the node belongs to. Since a node
is either occupied or unoccupied, $n(v_g)\in \{0,1\}$, we have
\begin{eqnarray}
  \label{eq:expectlaw}
  \langle n(v_g) \rangle & = & \sum_{n(v_g)=0,1} n(v_g) \Prob(n(v_g))
  \nonumber \\* & = & \Prob(n(v_g)\!=\!1) \equiv n_g\,.
\end{eqnarray}
We consider the evolution of a given set of mean occupations
$\bm{n}=(n_1, \dots , n_{d_M\!+\!1})$ to a new set $\bm{n}'$ induced
by the update algorithm
\begin{equation} \bm{n}' = \bm{f}(\bm{n}) \,,
  \label{eq:upd}
\end{equation}
where $\bm{f}=(f_1(\bm{n}),\dots ,f_{d_M\!+\!1}(\bm{n}))\,.$ The fixed
point $\bm{n}^\star$ of Eq.~(\ref{eq:upd}) solves the self-consistency
condition that we require for the stationary state of our model
system
\begin{equation} \bm{n}^\star =
  \bm{f}(\bm{n}^\star) \,. \label{eq:scc}
\end{equation}

Having obtained the fixed points $\bm{n}^\star$, we can compute the mean life time,
see Eq.~(\ref{eq:tau_g}) below, and the number of occupied neighbors by
\begin{equation}
  \langle n(\dvg) \rangle = \sum_{l=1}^{d_M+1} \Lgl n_l\,.
\end{equation}

\subsection{Update map}

\subsubsection{General approach}

The update map $\bm{f}(\bm{n})$ is constructed following the steps of the update algorithm, random influx and application of the deterministic window rule.  To be specific, we consider a node $v_g \in S_g$, with $g=1, \dots , d_M\!+\!1$, and its neighbourhood $\dvg$. Then the update algorithm says that an empty node $v_g$ is occupied by the influx with probability $p$. Next, the window condition is applied: An occupied node $v_g$ survives if the neighbourhood fulfills the window condition, otherwise it is emptied. We denote the occupation of a node $v_g$ after the influx by $n(\tildev_g)$ and the occupation of $v_g$ after the update by $n(v'_g)\,$. A microscopic configuration of $\dvg$ after the
influx is denoted by $\tildeC= C(\dvtg) =\left( n(\tildew) \right)_{ \tildew\in \dvtg}$, which is a list of occupations of all neighbors of $v_g$. Then we have
\begin{equation}
  \Prob(n(v'_g)) = \sum_{n(v_g)=0,1} \!\! T_{\text{update}}(n(v'_g)|n(v_g))
  \Prob(n(v_g)) \,, \label{eq:scexplicit}
\end{equation}
where 
\begin{eqnarray}
  \label{eq:transmat}
  T_{\text{update}}(n'|n) & = & (\delta_{n',1} \delta_{n,1} + p
  \delta_{n',1} \delta_{n,0}) \mathbb{W}(\dvtg) \nonumber \\* && +
  \delta_{n',0} \delta_{n,1} \left( 1-\mathbb{W}(\dvtg) \right) 
\end{eqnarray}
is the transition matrix of the update of $n(v_g)$ to $n(v_g')$ depending
on the window condition
\begin{equation}
  \label{eq:wc}
  \mathbb{W}(\dvtg) = \openone(t_L \leqslant n(\dvtg) \leqslant t_U)\,.
\end{equation}
Specifying Eq.~(\ref{eq:transmat}) to $n(v_g')=1$ we obtain from Eqs.
(\ref{eq:expectlaw}), (\ref{eq:scexplicit}), and (\ref{eq:transmat})
\begin{equation}
  \label{eq:scspec}
  n'_g = \left[ n_g + p(1-n_g) \right] \mathbb{W}(\dvtg)\,.
\end{equation}
Averaging over all configurations $\{ \tildeC \} $ leads to
\begin{equation}
  \label{eq:scav}
  n'_g = \left[ n_g + p(1-n_g) \right] \PW{g}\,,
\end{equation}
where $\PW{g} = \Prob(\mathbb{W}(\dvtg)\!=\!1)$ is the probability that after the influx the
neighbours of $v_g$ fulfill the window condition, 
\begin{eqnarray}
  \label{eq:pw}
  \PW{g} & = & \left\langle \mathbb{W}(\dvtg)
  \right\rangle_{\{\tildeC\}} = \sum_{\{\tildeC\}}
  \mathbb{W}(\dvtg) \Prob(\tildeC) \nonumber \\* 
  & = &  \sum_{\{\tildeC\}}
  \mathbb{W}(\dvtg) \sum_{\{C\}} T_{\text{influx}}(\tildeC|C)
  \Prob(C) \,.
\end{eqnarray}
In the next subsections \ref{sec:microstate} and \ref{sec:randominflux} we
determine the probability $\Prob(C)$ of a microstate $C=C(\dvg)$ and
the transition probability $T_{\text{influx}}(\tildeC|C)$ for the
transition from $C$ to $\tildeC$ induced by the influx. In subsection
\ref{sec:windowrule} $\PW{g}$ is explicitly determined.

\subsubsection{Probability of a microstate \label{sec:microstate}}

\begin{figure}[t]
  \centering \psfrag{dv}{$\dvg$} \psfrag{v}{$v_g$}
  \psfrag{S1}{$(\dvg)_1$} \psfrag{S2}{$(\dvg)_2$}
  \psfrag{S3}{$(\dvg)_3$} \psfrag{S4}{$(\dvg)_4$}
\includegraphics[width=0.5\columnwidth]{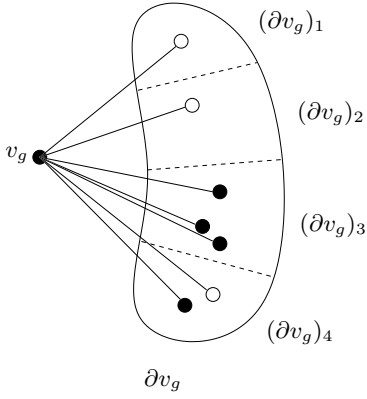}
\caption{A diagram of a microscopic configuration of $v_g$'s
  neighborhood.  Filled and empty circles represent occupied nodes and
  unoccupied nodes, respectively.}
\label{fig:dv}
\end{figure}

We introduce the notation $(\dvg)_l = \dvg \cap S_l$ for the set of
neighbors of $v_g$ belonging to group $S_l$, $|(\dvg)_l|=\Lgl$. A
microscopic configuration of $(\dvg)_l$ is denoted by $C_l =
(n(w))_{w \in (\dvg)_l}$. Figure~\ref{fig:dv} shows an
example configuration. Furthermore, a microscopic configuration $C_l$
with $k_l=n((\dvg)_l)$ occupied nodes is denoted by $C_{l|k_l}$. Such a
configuration has probability
\begin{equation}
  \label{eq:probcl}
  \Prob(C_{l|k_l}) = (n_l)^{k_l} (1-n_l)^{\Lgl-k_l} \,.
\end{equation}
There are ${L_{gl} \choose k_l}$ equivalent microconfigurations with the
same number of occupied nodes $k_l$. Multiplying this number yields, of course, the binomial distribution. The multiplicity has to be taken into account when calculating the average occupation of $(\dvg)_l$
\begin{equation}
  \label{eq:canonical}
  \langle n((\dvg)_l)\rangle_{\{C_{l|k_l}\}} = \sum_{k_l=0}^{L_{gl}} k_l
  {L_{gl} \choose k_l} \Prob(C_{l|k_l}) = L_{gl} n_l\,.
\end{equation}

The probability of a microscopic configuration $C_{\{k_l\}} = \bigcup_l
C_{l|k_l}$ of the whole neighbourhood $\dvg$ is determined by the
occupation numbers $\{k_l\}\,,\ l=1, \dots, d_M\!+\!1$. It factorises as
\begin{equation}
  \label{eq:probckl}
  \Prob(C_{\{k_l\}}) = \prod_{l=1}^{d_M\!+\!1} \Prob(C_{l|k_l})\,.
\end{equation}

\subsubsection{Random influx\label{sec:randominflux}}

The probability of the transition between two microstates $C_{l|k_l}$ and
$\tildeC_{l|k_l}$ is
\begin{equation}
  \label{eq:tinflux}
  T_{\text{influx}} (\tildeC_{l|k_l}|C_{l|k_l}) =
  p^{\tildekl-k_l} (1-p)^{\Lgl-\tildekl} \openone (\tildekl-k_l \geqslant
  0) \,. 
\end{equation}
Here, $\tildekl-k_l$ is the number of empty nodes in $C_{l|k_l}$
which become occupied, $\Lgl-\tildekl$ is the number of nodes of
$C_{l|k_l}$ which remain empty. The factor $\openone(\tildekl-k_l
\geqslant 0)$ is introduced the number of occupied nodes can not
decrease during the influx. There are ${\Lgl-k_l \choose
  \tildekl-k_l}$ different microconfigurations $\tildeC_{l|k_l}$
which can be reached from one microconfiguration $C_{l|k_l}$ with the
same probability, cf. Eq.~(\ref{eq:tinflux}). Again, there are ${\Lgl \choose
  k_l}$ equivalent microconfigurations $C_{l|k_l}\,$. Considering the
whole neighbourhood we have
\begin{equation}
  \label{eq:ticomplete}
  T_{\text{influx}} (\tildeC_{\{k_l\}}|C_{\{k_l\}}) =
  \prod_{l=1}^{d_M\!+\!1} T_{\text{influx}} (\tildeC_{l|k_l} |
  C_{l|k_l}) \,.  
\end{equation}

\subsubsection{Window rule\label{sec:windowrule}}

After the random influx which leads to a transition from $C = C(\dvg)$
to $\tildeC = C(\dvtg)$ it is tested whether or not the window
condition Eq.~(\ref{eq:wc}) $\mathbb{W}(\dvtg) = \openone(t_L\!
\leqslant\! n(\dvtg)\!  \leqslant\! t_U)$ is fulfilled. The
probability that the window condition is fulfilled is given by
Eq.~(\ref{eq:pw}). Replacing the sums over all microstates of $C$ and
$\tildeC$ by the sums over all mesostates $C_{\{k_l\}}$ and
$\tildeC_{\{k_l\}}$ with occupation numbers $\{k_l\}$ and $\{\tildekl\}$ we
have to account for the multiplicities derived above. This leads to
\begin{eqnarray}
  \label{eq:pwfinal}
  \PW{g} & = & \left[\,\sum_{\tildekl=0}^{\Lgl} \,\right]
  _{l=1}^{d_M\!+\!1} \!\! \openone \left( t_L \leqslant n(\dvtg) \leqslant
    t_U \right) \nonumber \\*
  && \times \prod _{l=1}^{d_M\!+\!1} \sum _{k_l=0}^{\Lgl} {\Lgl-k_l \choose
    \tildekl-k_l} {\Lgl \choose k_l} \\*
  && \times T_{\text{influx}}
  (\tildeC_{l|k_l} | C_{l|k_l}) \Prob(C_{l|k_l}) \,, \nonumber
\end{eqnarray}
where $n(\dvtg) = \sum_{l=1}^{d_M+1} \tildekl$. The factors
$T_{\text{influx}}(\tildeC_{l|k_l} | C_{l|k_l})$ and $\Prob(C_{l|k_l})$ are given by
Eqs. (\ref{eq:tinflux}) and (\ref{eq:probcl}). Since by definition
${\Lgl - k_l \choose \tildekl - k_l} \equiv 0$ for $\tildekl \!  -\!
k_l < 0$ and $\tildekl \!- \!k_l > \Lgl\! - \!  k_l\,,$ it is not
necessary to explicitly write the factor $\openone(\tildekl-k_l \geqslant 0)$, in contrast to
Eq.~(\ref{eq:tinflux}). We obtain
\begin{eqnarray}
  \label{eq:pwfinexpl}
  \PW{g} & = &\left[\, \sum_{\tildekl=0}^{\Lgl} \,\right]
  _{l=1}^{d_M\!+\!1} \!\! \openone \left( t_L \leqslant \sum_{l=1}^{d_M\!+\!1} \tildekl
    \leqslant t_U \right) \nonumber \\*
  && \times \prod _{l=1}^{d_M\!+\!1} \sum _{k_l=0}^{\tildekl} {\Lgl-k_l \choose
    \tildekl-k_l} {\Lgl \choose k_l} \\* 
  && \times p^{\tildekl-k_l} (1-p)^{\Lgl-\tildekl} n_l^{k_l}
  (1-n_l)^{\Lgl-k_l}\,, \nonumber 
\end{eqnarray}
which holds for all groups $g=1, \dots, d_M\!+\!1$. We can simplify
Eq.~(\ref{eq:pwfinexpl}) observing
\begin{equation}
  \label{eq:simplify1}
  {\Lgl \choose k_l} {\Lgl - k_l \choose \tildekl - k_l} = {\Lgl
    \choose \tildekl} {\tildekl \choose k_l}  \,.
\end{equation}
Applying the binomial formula
\begin{equation}
  \label{eq:simplify2}
  \sum_{k_l=0}^{\tildekl} {\tildekl \choose k_l} n_l^{k_l}
  (1-n_l)^{\tildekl-k_l} p^{\tildekl-k_l} = \left[ n_l + (1-n_l)p
  \right]^{\tildekl} \,,
\end{equation}
we can carry out the sum over the $k_l$ in Eq.~(\ref{eq:pwfinexpl}) and
obtain
\begin{eqnarray}
  \label{eq:pwfinsimp}
  \PW{g} & = & \left[\, \sum_{\tildekl=0}^{\Lgl} \,\right]
  _{l=1}^{d_M\!+\!1} \!\! \openone \left(t_L \leqslant \sum_{l=1}^{d_M\!+\!1} \tildekl
    \leqslant t_U\right) \nonumber \\* 
  && \times \prod _{l=1}^{d_M\!+\!1} {\Lgl \choose  \tildekl}
  \left[ n_l+(1-n_l)p \right] ^{\tildekl} \nonumber \\* 
  && \times [1-n_l-(1-n_l)p]^{\Lgl-\tildekl} \,.
\end{eqnarray}
Since the mean occupation of some node in $S_l$ after the influx is
$n_l+(1-n_l)p = \tilden_l\,$, Eq.~(\ref{eq:pwfinsimp}) can be written in
a compact way as
\begin{eqnarray}
  \label{eq:pwfinsimp2}
  \PW{g} & = & \left[\, \sum_{\tildekl=0}^{\Lgl} \,\right]
  _{l=1}^{d_M\!+\!1} \!\! \openone \left(t_L \leqslant \sum_{l=1}^{d_M\!+\!1} \tildekl
    \leqslant t_U\right) \nonumber \\* 
  && \times \prod _{l=1}^{d_M\!+\!1} {\Lgl \choose  \tildekl}
    \tilden_l ^{\tildekl} (1-\tilden_l)^{\Lgl-\tildekl} \,,
\end{eqnarray}
which has an obvious intuitive meaning: The probability that
$\tildekl$ of the $\Lgl$ nodes of $(\dvtg)_l$ are occupied is ${\Lgl
  \choose \tildekl} \tilden_l^{\tildekl} (1-\tilden_l)^{\Lgl-\tildekl}
\,$ and $\PW{g}$ is obtained by multiplying over all groups and summing
over all occupations obeying the window condition. Thus, all
quantities in Eq.~(\ref{eq:scav}) are determined and we are able to
calculate the mean occupation of all groups $\bm{n}$, e.g. by
iteration of Eq.~(\ref{eq:upd}).

\subsection{Mean life time\label{sec:meanlifetime}}

In a similar way, we are now able to calculate the mean life time
(expectation of life) of an occupied node $v_g$ in group $S_g$. Its
expectation value is defined as
\begin{equation}
  \label{eq:exptau1}
  \langle \tau(v_g) \rangle \equiv \tau_g = \sum_{\sigma=0}^{\infty}
  \sigma \Prob (\tau(v_g)=\sigma) \,, 
\end{equation}
where $\Prob (\tau(v_g)=\sigma)$ is the probability that an occupied
node $v_g \in S_g$ remains occupied in $\sigma$ subsequent steps and
disappears in the following step,
\begin{align}
  &\Prob(\tau(v_g)=\sigma) = \\ &\Prob (n_{\sigma\!+\!1}(v_g)\!=\!0,\,
  n_\sigma(v_g)\!=\!1,\, \dots ,\, n_1(v_g)\!=\!1 \, | \, n_0(v_g)\!=\!1)\,.
  \nonumber
\end{align}
This can be expressed in terms of the update transition matrix
$T_{\text{update}} (n'|n)$ introduced in Eq.~(\ref{eq:transmat}). We use
the shorthand $T^{(\sigma)}(1|1) = T_{\text{update}}
(n_\sigma(v_g)\!=\!1 \, | \, n_{\sigma\!-\!1}(v_g)\!=\!1)$ and
$T^{(\sigma)}(0|1)$ defined accordingly and write
\begin{align}
  \label{eq:ProbTau}
  & T^{(\sigma\!+\!1)}(0|1) T^{(\sigma)}(1|1) \dots T^{(1)}(1|1) \\* & =
  (1-\mathbb{W}(\dvtg^{(\sigma\!+\!1)})) \mathbb{W}(\dvtg^{(\sigma)})
  \dots \mathbb{W}(\dvtg^{(1)})\,, \nonumber
\end{align}
where $\dvtg^{(\sigma)}$ denotes the neighbourhood of $v_g$ after the
influx in the $\sigma$-th step of the iteration.

Now we take the average of Eq.~(\ref{eq:ProbTau}) over all possible
configurations $\tildeC^{(\sigma)}.$ Assuming that the
configurations in consecutive steps are independent, the average of
Eq.~(\ref{eq:ProbTau}) factorizes. In the steady state $\langle
\mathbb{W}(\dvtg^{(\sigma)}) \rangle _{\{\tildeC^{(\sigma)}\}} = \PW{g}$ is
independent of the time step $\sigma$ and given by
Eq.~(\ref{eq:pwfinsimp2}). This yields
\begin{equation}
\Prob(\tau(v_g)=\sigma)\, =\, (1-\PW{g}) (\PW{g})^\sigma\,.
\end{equation}
Observing $\sum_{\sigma=0}^{\infty} \sigma (\PW{g})^\sigma = \PW{g}/(1-\PW{g})^2$ we
finally obtain
\begin{equation}
  \label{eq:tau_g}
  \tau_g = \frac{\PW{g}}{1-\PW{g}} \,.
\end{equation}

\section{Simple Special Cases\label{sec:specialcases}}

In this section we discuss the numerical calculation of statistical
network properties and consider special cases which allow essential simplifications. 
The general case is considered in Sec.~\ref{sec:evaluation}.

We consider the model on the basegraph $G_d^{(m)}$ with parameters $[t_L, t_U]$ and $p$. We formulate the mean field theory for an architecture with module dimension $d_M$. The link matrix $\mathbb{L}=(L_{gl})$ is given by Eq.~(\ref{eq:linkmatrix}). The mean occupations of the groups in the steady state are the fixed points $\bm{n}^\star$ of Eq.~(\ref{eq:scc}). 
In general, Eq.~(\ref{eq:scc}) is a system of $d_M+1$ multivariate
polynomial equations in the form of Eq.~(\ref{eq:scav}). The factor
$\PW{g}$ is a polynomial in $p$ of maximal order $\kappa$, where $\kappa
= \sum_{i=0}^{m}{d \choose i}$ is the degree of a node on the base
graph. $\PW{g}$ is also a multivariate polynomial in $n_l$,
$l=0,\dots, d_M+1$. The maximal exponent of each $n_l$ is $\Lgl$, such
that the order of the multivariate polynomial is $\kappa$, because
$\sum_{l=1}^{d_M\!+\!1}\Lgl = \kappa$. Note, that it is not possible
to neglect higher orders of the polynomial, since the binomial
coefficients may be very large. That is, in general we have to solve very
complex equations.

In the following we treat groups of nodes for which Eq.~(\ref{eq:scc}) can be simplified in good approximation exploiting structural properties of the link matrix and properties of the groups of singletons and stable holes.

\subsection{Singletons\label{sec:singletons}}

Singletons are surrounded only by stable holes. They occur in most patterns, in static patterns as well as in dynamic ones. Therefore, singletons are important. Their treatment is simple, since they decouple in very good approximation from the rest.

\begin{figure}[t]
\centering
\includegraphics[width=\columnwidth]{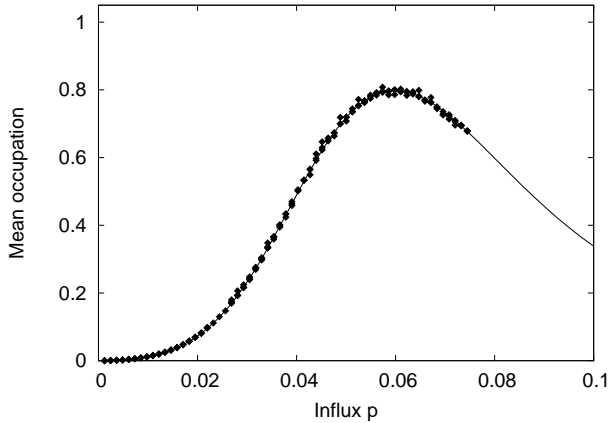}
\caption{Mean field results (line) for singletons on graph
  $G_{12}^{(2)}$ with $[t_L,t_U] = [1,10]$ compared with simulation
  (symbols). For small $p$, $n^\star_{\mathrm{sing}} = \kappa p^2$, for $p \gtrapprox 0.075$ the patterns in the simulation become {\em transient}, such that no group averages are
  available. Singletons occur in various patterns, cf. Fig.~\ref{fig:mfavssim}. \label{fig:mfsingletons}}
\end{figure}

Typically, stable holes are surrounded by more than $t_U$ occupied nodes, such that their occupation is zero after each update step. An occupied stable hole is a very rare event, and in good approximation we can assume that all stable holes are empty. Singletons are surrounded by $\kappa$ stable holes. The probability that $k$ out of $\kappa$ empty stable holes are occupied by the influx is ${\kappa \choose k} p^k (1\!-\!p)^{\kappa\!-\!k}$. Thus, the probability that the window condition is fulfilled is given by
\begin{equation}
  \PW{\rm{sing}} = \sum_{k=t_L}^{t_U} {\kappa \choose k} p^k (1-p)^{\kappa-k}\,.
\end{equation}
This follows also directly from Eq.~(\ref{eq:pwfinsimp}) for singletons setting all hole groups empty. Note, that $\PW{\rm{sing}}$ only depends on $p$, and $\PW{\rm{sing}} \approx \kappa p$ for small $p$. For singletons Eq.~(\ref{eq:scav}) has the unique solution
\begin{equation}
  \label{eq:n_sing} n^\star_{\rm{sing}} = \frac{p\PW{\rm{sing}}}{1-(1-p)\PW{\rm{sing}}}\,.
\end{equation}
The mean occupation $n^\star_{\mathrm{sing}}(p)$ is plotted in Fig.~\ref{fig:mfsingletons} and compared with simulation results.

\subsection{Occupied Core in Static Patterns\label{sec:occupiedcore}}

Self-coupled nodes in static patterns have only stable holes and
members of their own group as neighbors. They appear for instance in
$d_M=2$, 4, and 6 patterns, cf. \cite{STB}. It is
characteristic for these patterns, that the self-coupled nodes have a
high mean occupation and thus suppress the occupation of the stable
holes. Supposing a high occupation of the self-coupled nodes we can
consider stable holes as unoccupied. Then, instead of the system of Eqs.~(\ref{eq:scc})
we have only one independent equation for the respective occupied core group.

As an example we consider the $d_M=4$ pattern. $S_1$ are singletons, the group of
self-coupled nodes is $S_2$. The stable hole groups $S_3$, $S_4$, and
$S_5$ are suppressed by the occupation of $S_2$ if $n_2 \gg
n_2^{\rm{critical}} = t_U / \min (L_{3,2}, L_{4,2}, L_{5,2}) = 0.56$. With this constraint we
can simply consider the union $S_{\rm{sh}} = S_3 \cup
S_4 \cup S_5$ and put $n_3\!=\! n_4 \!=\! n_5 = n_{\rm{sh}} = 0$ for all groups of
stable holes. This simplifies the link matrix as given in Fig.~\ref{fig:matrixsimplification}. The nodes in $S_1$, singletons, are treated as given above, $n_1 = n_{\rm{sing}}^\star$. For the self-coupled nodes in $S_2$ we have a polynomial of 4th order in $n_2$ and of 80th order in $p$,
\begin{align}
  \label{eq:onegroup_8c}
  n_2' = & (n_2 + p (1\!-\!n_2)) \sum_{k_2=0}^3 \sum_{k_{\rm{sh}}=0}^{76}
  \openone (1\leqslant k_2\!+\!k_{\rm{sh}} \leqslant 10) \nonumber \\
  & \times {3 \choose k_2} (n_2 + p (1\!-\!n_2))^{k_2} (1 - n_2 - p
  (1\!-\!n_2))^{3\!-\!k_2} \nonumber \\
  & \times {76 \choose k_{\rm{sh}}} p^{k_{\rm{sh}}} (1 -
  p)^{76\!-\!k_{\rm{sh}}} \,.
\end{align}

\begin{figure}[t]
\begin{displaymath}
  \renewcommand{\arraystretch}{1.3}
  \begin{array}{|l|ccccc|}
    \hline
    & S_{1} & S_{2} & S_{3} & S_{4} & S_{5} \\
    \hline
    v_{1} & & & 6 & 36 & 37 \\
    v_{2} & & 3 & 27 & 40 & 9 \\
    v_{3} & 1 & 18 & 41 & 18 & 1 \\
    v_{4} & 9 & 40 & 27 & 3 & \\
    v_{5} & 37 & 36 & 6 & & \\
    \hline
  \end{array}
  \quad\longrightarrow\quad
  \begin{array}{|l|ccc|}
    \hline
    & S_{1} & S_{2} & S_{\rm{sh}} \\
    \hline
    v_{1} & & & 79 \\
    v_{2} & & 3 & 76 \\
    \hline
  \end{array}
\end{displaymath}
\caption{The link matrix can be simplified if the stable holes are practically always empty after the update, i.e. for $n_2\gg n_2^{\rm critical}$, see text. Then we lump all stable hole groups $S_3$, $S_4$ and $S_5$ to $S_{\rm sh}$.\label{fig:matrixsimplification}}
\end{figure}

\begin{figure}[t]
\centering
\includegraphics[width=\columnwidth]{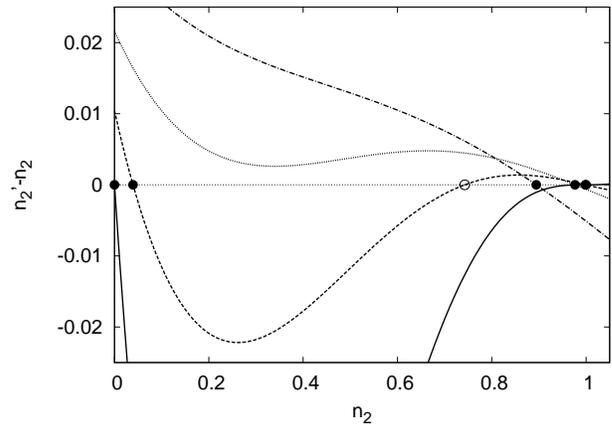}
\caption{Fixed points for self-coupled nodes in the $d_M=4$ pattern on graph $G_{12}^{(2)}$ with $[t_L,t_U] = [1,10]$. The lines give the function $n_2'\!-\!n_2$ from the mean field theory for $p=0$ ({\em solid}), $p=0.015$  ({\em dashed}), $p=0.025$ ({\em dotted}), and $p=0.035$ ({\em dash-dotted}). The symbols mark stable ($\bullet$) and unstable ($\circ$) fixed points. Note, that the fixed points $n_2^\star=0$ and $n_2^\star=0.039$ are much smaller than $n_2^{\rm critical}$, which is a range where the equation is actually not valid.\label{fig:mfS2dM4}}
\end{figure}

In Fig.~\ref{fig:mfS2dM4} we plotted $n_2'\!-\!n_2$ over $n_2$ for representative choices
of the influx parameter $p$. For small values of $p$, $n_2'$ is very close to $n_2$ such that the common way to plot $n_2'$ over $n_2$ is inconvenient. The roots of $n_2'\!-\!n_2=0$ give the fixed points $n_2^\star$ of Eq.~(\ref{eq:onegroup_8c}). Fixed points with $|{\rm d}/{\rm d}n_2\, n_2'| < 1$, i.e. with
\begin{equation}
  -2 < \frac{\rm d}{{\rm d}n_2} (n_2'\!-\!n_2) < 0 \,,
\end{equation}
are stable.

\begin{figure}[t]
\centering
\includegraphics[width=\columnwidth]{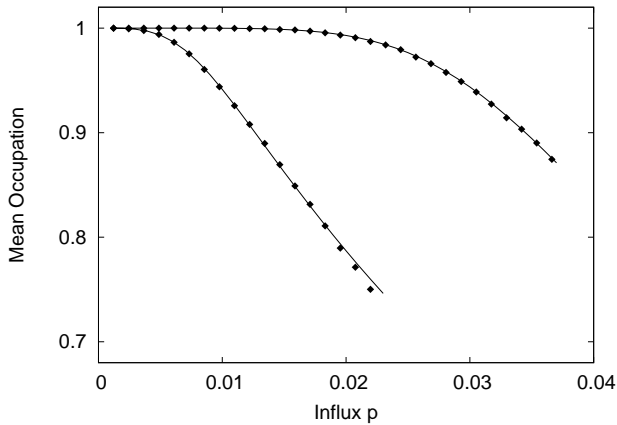}
\caption{Mean field results for nodes of $S_2$ in a $d_M=4$ pattern ({\em upper curve}) and nodes of group $S_3$ in a $d_M=6$ pattern ({\em lower curve}), both on graph $G_{12}^{(2)}$ with $[t_L,t_U] = [1,10]$ compared with the respective values from simulations ({\em symbols}). Lines end at the value of $p$ where the patterns become unstable in the simulations, cf. Fig.~\ref{fig:mfavssim} ({\em right}).} \label{fig:mfonegroup}
\end{figure}

The case of $d_M=6$ is treated in the same way. For both $d_M=4$ and $6$ the stable fixed points are in very good agreement with the simulations, cf. Fig.~\ref{fig:mfonegroup}. 

The case of $d_M=2$ can not be treated this way, however. It is easy to see that for $p=0$ the respective iteration equation for the occupied group $S_1$ is $n_1'=n_1^2$, which has the fixed points $n_1^\star=0$ and $1$. The fixed point $n_1^\star=1$ corresponds to the perfect 2-cluster pattern but it is unstable. Generically, the iteration converges to the stable fixed point $n_1^\star=0$. Also for $p>0$ the iteration will not produce the correct result of the simulations, even if we consider the full set of equations for all groups, cf. Eq.~(\ref{eq:scc}). The reason is simple: in the 2-cluster pattern an occupied node has only one occupied neighbor, which obviously obstructs the mean field approach. Correlations between the two occupied neighbors are not negligible, they are treated in the next section.

\section{2-Cluster Pattern with Correlations\label{sec:correlation}}

\subsection{Mean occupation}

\begin{figure}[h]
\centering
\psfrag{dv}{$\partial v$} \psfrag{dw}{$\partial w$} \psfrag{v}{$v$}
\psfrag{w}{$w$} 
\includegraphics[width=0.5\columnwidth]{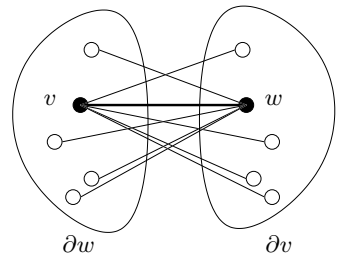}
\caption{Nodes $v$ and $w$ from $S_1$ forming a 2-cluster and their disjoint neighborhoods $\partial v$ and $\partial w$. The line width of the links corresponds to the correlation strength. In the ideal pattern $v$ and $w$ are occupied and the other nodes are empty.
\label{fig:correlation}}
\end{figure}

In any 2-cluster pattern the occupied nodes form pairs of mutually
stimulating nodes, cf. Fig.~\ref{fig:correlation}. Their survival
significantly depends on the presence of the respective partner, they are
strongly correlated. This holds for all 2-cluster patterns on base
graphs with arbitrary numbers of allowed mismatches $m$. The pattern
module of a 2-cluster pattern on $G_d^{(m)}$ is of dimension $d_M=m$
and the group of 2-cluster nodes shall be denoted by $S_1$ in all
these cases.

We measured in simulations the correlation between nearest neighbors in different
patterns. The correlation is quantified by the two-point connected
correlation function
\begin{equation}
  G_c(v,w) = \langle n(v) n(w) \rangle - \langle n(v) \rangle \langle n(w)
  \rangle \,.
\end{equation}
If $G_c(v,w) > 0$, $v$ and $w$ are correlated, if $G_c(v,w) = 0$, they are uncorrelated, and if $G_c(v,w) < 0$, they are anti-correlated. We determined $G_c(v_i,w_j)$ for $v_i \in S_i$ and $w_j \in (\partial v_i)_j = \partial v_i \cap S_j$ a nearest neighbor of $v_i$ and an element of $S_j$. The average over all members of $S_i$ and their neighbors in $S_j$ is denoted by $G_{ij} = \langle \langle G_c(v_i, w_j) \rangle _{v_i \in S_i} \rangle _{w_j \in (\partial   v_i)_j}$.  Tables~\ref{tab:correlation1} and \ref{tab:correlation2} show $G_{ij}$ measured for all $i$ and $j$ for the 2-cluster and the
8-cluster pattern, respectively. The correlation in the 2-cluster is the strongest and thus indeed relevant. Assuming independence gives qualitatively false results, as explained above. 

\begin{table}[h]
  \caption{$G_{ij}$ for the 2-cluster pattern, $p=0.0366$. The
    corresponding value for $G_{11}$ calculated in the mean field approach
    below is $\langle vw \rangle - \langle v \rangle^2 = 2.1\cdot 10^{-2}$.\\}
\label{tab:correlation1}
\begin{ruledtabular}
\begin{tabular}{clll}
  & \multicolumn{1}{c}{$S_1$} & \multicolumn{1}{c}{$S_2$} & \multicolumn{1}{c}{$S_3$} \\[0.5mm]
  \hline \\[-2.5mm]
  $S_1$ & $\phantom{-}2.2\cdot 10^{-2}$ & $-2.6\cdot 10^{-4}$ & $\phantom{-}5.4\cdot 10^{-20}$ \\
  $S_2$ & $-2.6\cdot 10^{-4}$ & $-3.1\cdot 10^{-6}$ & $\phantom{-}1.2\cdot 10^{-19}$ \\
  $S_3$ & $\phantom{-}5.4\cdot 10^{-20}$ & $\phantom{-}1.2\cdot 10^{-19}$ & $\phantom{-}0.0$ \\
\end{tabular}
\end{ruledtabular}
\end{table}

\newcommand{\st}{\scriptstyle} \newcommand{\sst}{\scriptscriptstyle}
\begin{table}[h]
  \caption{$G_{ij}$ for the 8-cluster pattern, $p=0.0366$. Entries
    are missing if there exist no nearest neighbors in these groups.\\} 
\label{tab:correlation2}
\begin{ruledtabular}
\begin{tabular}{clllll}
 & \multicolumn{1}{c}{$S_1$} & \multicolumn{1}{c}{$S_2$} & \multicolumn{1}{c}{$S_3$} & \multicolumn{1}{c}{$S_4$} & \multicolumn{1}{c}{$S_5$} \\[0.5mm]
\hline \\[-2.5mm]
$S_1$ & & & ${\st -}4.9\!\cdot\! 10^{\sst -6}$ & $\phantom{\st -}1.3\!\cdot\! 10^{\sst -23}$ & $\phantom{\st -}1.3\!\cdot\! 10^{\sst -23}$ \\
$S_2$ & & ${\st -}7.0\!\cdot\! 10^{\sst -3}$ & ${\st -}9.2\!\cdot\! 10^{\sst -6}$ & ${\st -}3.0\!\cdot\! 10^{\sst -22}$ & ${\st -}3.0\!\cdot\! 10^{\sst -22}$ \\
$S_3$ & ${\st -}4.9\!\cdot\! 10^{\sst -6}$ & ${\st -}9.2\!\cdot\! 10^{\sst -6}$ & ${\st -}5.0\!\cdot\!
10^{\sst -10}$ & $\phantom{\st -}0.0$ & $\phantom{\st -}0.0$ \\
$S_4$ & $\phantom{\st -}1.3\!\cdot\! 10^{\sst -23}$ & ${\st -}3.0\!\cdot\! 10^{\sst -22}$ & $\phantom{\st -}0.0$ & $\phantom{\st -}0.0$ & \\
$S_5$ & $\phantom{\st -}1.3\!\cdot\! 10^{\sst -23}$ & ${\st -}3.0\!\cdot\! 10^{\sst -22}$ & $\phantom{\st -}0.0$ & & \\
\end{tabular}
\end{ruledtabular}
\end{table}

Consequently, we must not factorize the joint probability $\Prob(n(v), n(w))$ for nodes $v$ and $w$ of a 2-cluster.
In the following we abbreviate $n(v)$ simply by $v$, and the occupation after the influx $n(\tildev)$ and after the update $n(v')$ by $\tildev$ and $v'$, respectively.
In general it holds $P(v)=\sum_{w} P(v, w)$. 
For symmetry reasons $\langle v \rangle = \langle w \rangle$ and also
$P(v, w)= P(w,v)$.  And, of course, there is the normalization
$\sum_{v, w} P(v, w) = 1$.

As a consequence of the strong correlation between nodes of a
2-cluster the update rule $\bm{n}'_{\text{cor}} \! = \!
\bm{f}_{\text{cor}} (\bm{n}_{\text{cor}}) $ is not only a function
of $n_1 \equiv \langle v \rangle = P(v\!=\!1)$, but also of the pair
correlation $\langle vw \rangle$.
\begin{equation}
  \bm{n}_{\text{cor}} = (\langle v \rangle, \langle vw \rangle , n_2, n_3)\,.
\end{equation}

We now construct the update map
$\bm{f}_{\text{cor}}$. Therefore, we once more determine the
transition matrix, cf. Eq.~(\ref{eq:transmat}), this time for the pair
$v$ and $w$
\begin{equation}
  P(v',w') = \sum _{v, w} T_{\text{update}} (v', w'|v, w) P(v, w)\,.
\end{equation}
It is constructed from the two subsequent steps of influx
and application of the window rule
\begin{align}
  \label{eq:TupdateCorr}
  T_{\text{update}} & (v', w'|v, w) = \nonumber \\ & \sum_{\tilde{v}, \tilde{w}}
  T_{\text{window}} (v', w'|\tilde{v}, \tilde{w})
  T_{\text{influx}} (\tilde{v}, \tilde{w}|v, w)\,.
\end{align}
The influx step, which is still independent for the nodes $v$ and $w$
is defined as
\begin{align}
  \label{eq:TinfluxCorr}
  T_{\text{influx}} ( & \tilde{v}, \tilde{w}| v, w) =
  T_{\text{influx}} (\tilde{v}|v) T_{\text{influx}} (\tilde{w}|w)
  = \nonumber \\ & (\delta_{\tilde{v}, 1} \delta_{v, 1} +
  \delta_{\tilde{v}, 1} \delta_{v, 0} p + \delta_{\tilde{v}, 0}
  \delta_{v, 0} (1-p)) \\ & \times (\delta_{\tilde{w}, 1} \delta_{w,
    1} + \delta_{\tilde{w}, 1} \delta_{w, 0} p + \delta_{\tilde{w}, 0}
  \delta_{w, 0} (1-p)) \,. \nonumber
\end{align}
%
It is now straightforward to specify $\tildeP (v, w) \equiv
P(\tilde{v}, \tilde{w}) = \sum _{v,w} T_{\text{influx}} (\tilde{v},
\tilde{w}| v, w) P(v, w)$ to 
\begin{eqnarray}
\tildeP (0,0) & = & P(0,0)(1-p)^2 \,, \nonumber \\
\tildeP (0,1) & = & P(0,0)p(1-p) + P(0,1)(1-p) \,, \nonumber \\
\tildeP (1,1) & = & P(0,0)p^2 + 2P(0,1)p + P(1,1) \label{eq:tildePcorr}\,, 
\end{eqnarray}
where we used $P(1,0) = P(0,1)$. 
Similarily, for the window rule we have
\begin{align}
  \label{eq:TwindowCorr}
  T & {}_{\text{window}} (v', w' | \tilde{v}, \tilde{w}) = \nonumber \\
  & (\delta_{v'\!, 1} \delta_{\tilde{v}, 1} \mathbb{W}({\partial \tilde{v}}|
  \tilde{w}) + \delta_{v'\!, 0} \delta_{\tilde{v}, 1}
  (1\!-\!\mathbb{W}({\partial\tilde{v}} | \tilde{w})) + \delta_{v'\!, 0}
  \delta_{\tilde{v}, 0}) \nonumber \\ 
  & \times (\delta_{w'\!, 1}
  \delta_{\tilde{w}, 1} \mathbb{W}({\partial\tilde{v}} | \tilde{w}) +
  \delta_{w'\!, 0} \delta_{\tilde{w}, 1} (1\!-\!\mathbb{W}({\partial\tilde{w}}
  | \tilde{v})) \nonumber \\
  & \phantom{\times ()} + \delta_{w'\!, 0} \delta_{\tilde{w}, 0}) \,,
\end{align}
where 
\begin{equation}
  \mathbb{W} (\partial \tilde{v}\, |\, \tilde{w} \! = \! n ) =
  \openone \left( t_L \leqslant n + \sum _{l=2} ^{3} \tildekl
    \leqslant t_U \right) \,. \label{eq:WindowCondCor}
\end{equation}
Here, we sum over the groups $S_2$ and $S_3$ and the occupation of the partner node $\tilde{w}$ in $S_1$ is explicitly taken into account. Note also the explicit dependence of the two factors in Eq.~(\ref{eq:TwindowCorr}) on $\tilde{v}$ and $\tilde{w}$.

Application of the window rule leads to
\begin{eqnarray}
  P' (0,0) & = & \tildeP(0,0) + \tildeP(0,1)
  (1\!-\!\mathbb{W}({\partial\tilde{w}} | 0)) \nonumber \\ && 
  + \tildeP(1, 0) (1\!-\!\mathbb{W}({\partial\tilde{v}} | 0)) \nonumber \\ &&  
  + \tildeP(1,1) (1\!-\!\mathbb{W}({\partial\tilde{w}}|1) ) (
  1\!-\!\mathbb{W}({\partial\tilde{v}}|1) )  \,, \nonumber \\ 
  P' (0,1) & = & \tildeP(0,1) \mathbb{W}({\partial\tilde{w}}|0)
  \nonumber \\ && 
  + \tildeP(1,1) (1\!-\!\mathbb{W}({\partial\tilde{v}}|1) )
  \mathbb{W}({\partial\tilde{w}}|1) \,, \nonumber \\  
  P' (1,0) & = & \tildeP(1,0) \mathbb{W}({\partial\tilde{v}}|0)
  \nonumber \\ && 
  + \tildeP(1,1) \mathbb{W}({\partial\tilde{v}}|1)
  (1\!-\!\mathbb{W}({\partial\tilde{w}}|1) ) \,, \nonumber \\  
  P' (1,1) & = & \tildeP(1,1) \mathbb{W}({\partial\tilde{v}}|1)
  \mathbb{W}({\partial\tilde{w}}|1) \label{eq:Ppcorr}\,.
\end{eqnarray}
Averaging Eq.~(\ref{eq:WindowCondCor}) over the neighborhood
microconfigurations $C(\partial \tilde{v} \backslash \tilde{w})$ of the
2-cluster nodes yields 
\begin{equation} 
  \langle \mathbb{W}({\partial\tilde{v}}|n) \rangle
  _{\{C(\partial \tilde{v} \backslash \tilde{w})\}} = \PW{1|n}\,, \label{eq:averagingW}
\end{equation} 
the conditional probability that the window condition is fulfilled for
$\tilde{v}$ given that the partner node $\tilde{w}$ has occupation
$n$. Explicitly, in analogy to Eq.~(\ref{eq:pwfinsimp2})
\begin{eqnarray}
  \label{eq:pwcorr}
  \PW{1|n} & = & \left[\, \sum_{\tildekl=0}^{L_{1l}} \,\right]
  _{l=2}^3 \!\! \openone \left(t_L  \leqslant n +
    \sum_{l=2}^3 \tildekl
    \leqslant t_U \right) \nonumber \\* 
  && \times \prod _{l=2}^3 {L_{1l} \choose  \tildekl}
  \tilden_l^{\tildekl} (1-\tilden_l)^{L_{1l}-\tildekl} \,.
\end{eqnarray}
Of course, it holds that
\begin{equation}
\sum_{n=0,1} \!\!\PW{1|n} \Prob(\tilde{w}\!=\!n) = \PW{1|0} (1\!-\!\langle
\tilde{w} \rangle ) + \PW{1|1} \langle \tilde{w} \rangle =
P^{\mathbb{W}}_1\,. 
\end{equation}

The four pair probabilities $P(v, w)$ are not independent. They obey
the symmetry condition and are normalized. We can express them by
$\langle v \rangle$ and $\langle vw \rangle$ as
\begin{eqnarray}
  P(1, 1) & = & \sum_{v, w} v w P (v, w) = \langle vw \rangle \,,
  \nonumber\\ 
  P(1, 0) & = & \langle v(1\!-\!w) \rangle = \langle v \rangle -
  \langle vw \rangle \,, \nonumber \\ 
  P(0, 1) & = & \langle (1\!-\!v)w \rangle = \langle w \rangle -
  \langle vw \rangle \,, \nonumber \\ 
  P(0, 0) & = & \langle (1\!-\!v)(1\!-\!w) \rangle = 1 \! - \!
  \langle v \rangle \! - \! \langle w \rangle \! + \! \langle vw
  \rangle ,
\end{eqnarray}
where we used $\langle v \rangle \equiv P(v\!=\!1) = P(1, 1) + P(1,
0)$.  Thus, we can use $\langle v \rangle$ and $\langle vw \rangle$ as
independent variables. The update rules for them obtained from
Eqs. (\ref{eq:tildePcorr}), (\ref{eq:Ppcorr}), and
(\ref{eq:averagingW}) are
\begin{align}
  \label{eq:vUpdate}
  \langle v \rangle ' =& \left[ - \langle vw \rangle (1\!-\!p)^2 +
    \langle v \rangle(1\!-\!p)(1\!-\!2p) + p(1\!-\!p) \right] \PW{1|0} \nonumber \\* 
  & + \left[ \langle vw \rangle (1\!-\!p)^2 + \langle v
    \rangle 2 p (1\!-\!p) + p^2 \right] \PW{1|1} \,,
\end{align}
\vspace{-6mm}
\begin{equation}
  \label{eq:vwUpdate}
  \langle vw \rangle ' = \left[ \langle vw \rangle (1\!-\!p)^2 + \langle v
    \rangle 2 p (1\!-\!p) + p^2 \right] (\PW{1|1})^2 .
\end{equation}
Neglecting the pair correlation reproduces the simple mean field
equation, of course.

Finally, the mean occupation of all other groups in the 2-cluster
pattern are obtained from Eq.~(\ref{eq:scav}) using $n_1 \equiv \langle v
\rangle$ from Eq.~(\ref{eq:vUpdate}).

\subsection{Mean life time}

In order to calculate $\Prob(\tau(v_1)=\sigma)$,
cf. Eq.~(\ref{eq:exptau1}), we need the transition probablilities
$T^{(\sigma)}(v'|v) = P(v', v) / P(v)$ again.  However, for the case of
strongly correlated nodes we have to consider the nodes in a 2-cluster
$v$ and $w$ as a pair.
\begin{eqnarray}
  P(v',v) & = & \sum _{w, w'} P(v', w'; v, w)\,, \\
  P(v', w'; v, w) & = & T_{\text{update}} (v', w'|v, w) P(v, w)\,.
\end{eqnarray}
Thus, the transition probability corresponding to $T^{(\sigma)}(v'|v)$
in Eq.~(\ref{eq:ProbTau}) is now given by
\begin{equation}
  \label{eq:TcorrSigma}
  T_{\text{cor}}^{(\sigma)}(v'|v) = \sum_{w, w'}
  T_{\text{update}}(v_t, w_t | v_{t\!-\!1},
  w_{t\!-\!1}) P(v, w) / P(v) \,, 
\end{equation}
where $P(v) = \sum_{w} P(v, w)$.

Equation~(\ref{eq:TcorrSigma}) can be applied to formula
(\ref{eq:ProbTau}) in order to obtain $\Prob(\tau(v_1)=\sigma)$. Explicitly we
have
\begin{eqnarray}
  T_{\text{cor}}^{(\sigma)}(1|1) & = & \left\{ \left[ (1\!-\!p)
      \mathbb{W}(\partial \tildev |0) + p\mathbb{W}(\partial \tildev |1) \right]
    P(1, 0) \right. \nonumber \\* 
  & & \left. + \mathbb{W}(\partial \tildev |1) P(1, 1) \right\} / P(1) \,, \\ 
  T_{\text{cor}}^{(\sigma)}(0|1) & = & 1 - T_{\text{cor}}^{(\sigma)}(1|1) \,.
\end{eqnarray}

Using that the two neighbourhoods $\partial v \backslash w$ and
$\partial w \backslash v$ are disjoint and assuming that their
configurations after the influx $C^{(\sigma)}(\partial \tilde{v} \backslash \tilde{w})$ in
subsequent time steps $\sigma$ are independent, we
average over all these configurations. This replaces all
$\mathbb{W}(\partial \tildev |n)$ by $\PW{1|n}$ given by
Eq.~(\ref{eq:pwcorr}). Expressing the $P(v, w)$ by the independent
variables $\langle v \rangle$  and $\langle vw \rangle$ we obtain
\begin{align}
  \langle T_{\text{cor}} (1|1) \rangle & {}_{\{C^{(\sigma)}(\partial \tilde{v} \backslash \tilde{w})\}} 
  = [(1\!-\!p)
  \PW{1|0} + p\PW{1|1}] \\*
  & + [(1\!-\!p)
  \PW{1|0} + (1\!-\!p)\PW{1|1}] \langle vw
  \rangle / \langle v \rangle \,. \nonumber 
\end{align}
This is independent of $\sigma$ in the steady state and we shortly write $\langle T_{\text{cor}} (1|1) \rangle$.
In the same way as in the uncorrelated case this yields
\begin{eqnarray}
  \langle \tau (v_1) \rangle = \tau_1 & = & \sum_{\sigma=0}^\infty \sigma \langle
  T_{\text{cor}} (0|1)\rangle 
  \langle T_{\text{cor}} (1|1) \rangle ^\sigma
  \nonumber \\ & = & \frac 
  {\langle T_{\text{cor}} (1|1) \rangle}{1-\langle
    T_{\text{cor}} (1|1) \rangle} \,.
\end{eqnarray}

Table~\ref{tab:groups2c} compares simulation and mean field results of
the mean occupation and the mean life time. They are all in very good
agreement.

\begin{table}[h]
\caption{Simulation vs. mean field results for groups in the 2-cluster pattern
  for $p\!=\!0.025$\label{tab:groups2c}\\} 
\begin{ruledtabular}
\begin{tabular}{lcccc}
& & $S_1$ & $S_2$ & $S_3$ \\[0.5mm]
\hline \\[-2.5mm]
mean occupation & $\langle \overline{n}(v) \rangle_{S_i}$ &
0.993 & 0.0004 & 0.000 \\
& $n_{\text{MFA}}$ &
0.993 & 0.0003 & 0.000 \\[1mm]
mean life time & $\langle \overline{\tau} (v) \rangle_{S_i}$ &
6115 & 0.017 & 0.000 \\
& $\tau_{\text{MFA}}$ &
6378 & 0.014 & 0.000 \\[1mm]
occ. neighbours & $\langle \overline{n}(\partial v) \rangle_{S_i}$ &
1.002 & 10.94 & 55.60 \\
& $n(\partial v)_{\text{MFA}}$ &
1.001 & 10.94 & 55.62 \\
\end{tabular}
\end{ruledtabular}
\end{table}

\section{Evaluation of the General Case\label{sec:evaluation}}

We notice that the solution of Eq.~(\ref{eq:scc}) can be very
laborious, apart from very few simple special cases. In general we can
determine the stable fixed points $\bm{n}^\star$ by iteration starting
with some reasonable initial values $\bm{n}^0$. 

The suitable choice of the initial values is crucial, because
Eq.~(\ref{eq:scc}) often has multiple stable fixed points. However, the
basin of attraction is rather large for fixed points corresponding to
dynamic patterns.

{\em Generic} initial values, i.e. values near the stationary results
of the respective simulation, converge to the simulation results.

For a nongeneric choice of $\bm{n}^0$ the iteration may lead to
results which do not correspond to the behaviour of the simulated
system. For example, {\em homogenous} initial values $n^0_g=n^0$ for all
groups lead to a homogenous fixed point $n^\star_g=n^\star$. This is
equivalent to an architecture with only one group. An analysis of one
single polynomial equation like in subsection
\ref{sec:occupiedcore} can be possible in this case. \cite{BBC89}
analyzes a similar situation. {\em Symmetric} initial conditions $\bm{n}^0$ with
$n^0_g=n^0_{d_M\!+\!2\!-\!g}$ lead to symmetric fixed points. This is
due to the symmetry of the link matrix
\begin{equation}
  \label{eq:Lijsymmetry}
  L_{gl} = L_{d_M+2-g, d_M+2-l} \,,
\end{equation}
which is inherited to the update map $\bm{f}(\bm{n})$.

For generic initial conditions we computed the fixed points for
various pattern modules and an interesting interval of the parameter
$p$ on $G_{12}^{(2)}$ and $[t_L, t_U]=[1,10]$ and compare the results with simulation data.

In the simulations we obtained the mean occupation as
\begin{equation}
  \overline{n}(v) = \frac{1}{T_1-T_0} \sum_{t\in (T_0, T_1]} n_t(v) \,,
\end{equation}
the mean number of occupied neighbors of $v$
\begin{equation}
  \overline{n}(\partial v) = \frac{1}{T_1-T_0} \sum_{t\in (T_0, T_1]} n_t(\partial v) \,,
\end{equation}
and the mean life time
\begin{equation}
  \overline{\tau}(v) = \frac{1}{b(v)+n_{T_0}} \sum_{t\in (T_0, T_1]} n_t(v)\,,
\end{equation}
where $b(v)$ is the number of births during the observation time,
i.e. the number of new occupations of the node by the influx. Of
course, $b(v)+n_{T_0} \neq 0$ must be fulfilled, otherwise
$\overline{\tau}(v)$ has no meaning. 

\begin{figure*}[p]
\centering
\includegraphics[width=1.0\columnwidth]{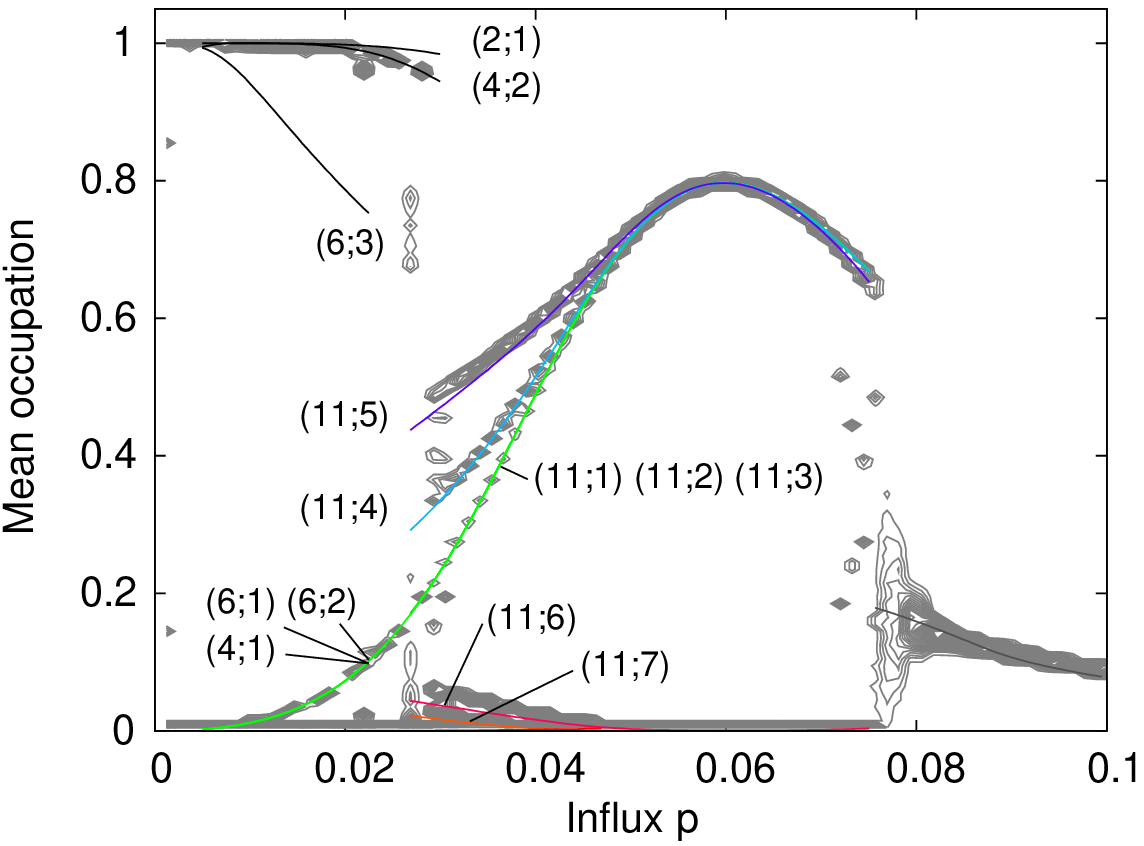}
\includegraphics[width=1.0\columnwidth]{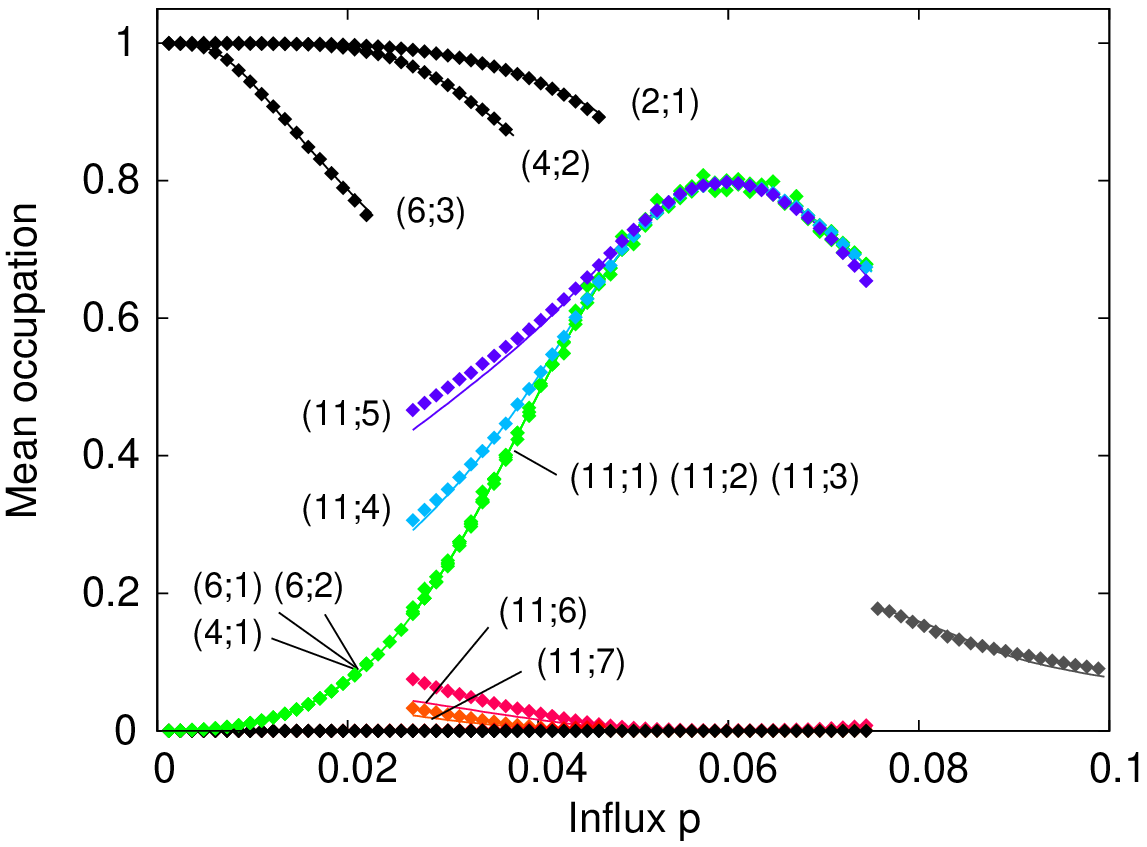}
\includegraphics[width=1.0\columnwidth]{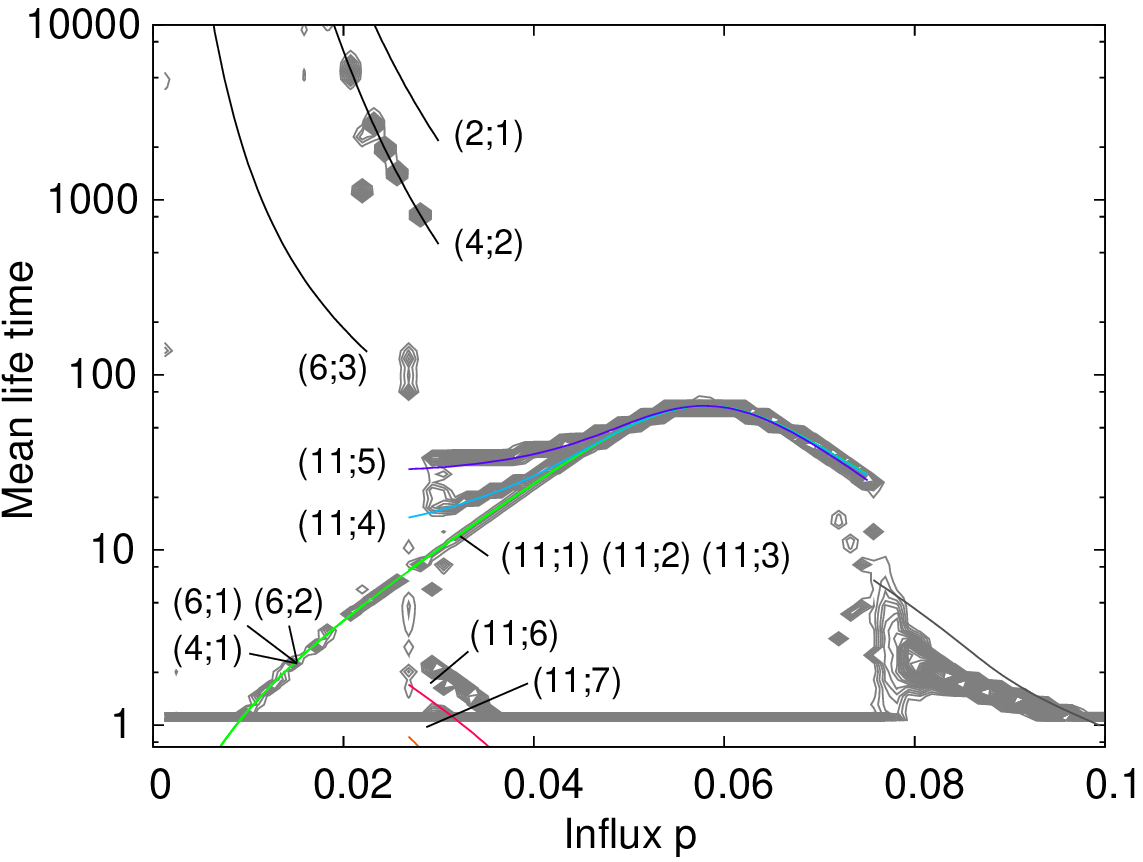}
\includegraphics[width=1.0\columnwidth]{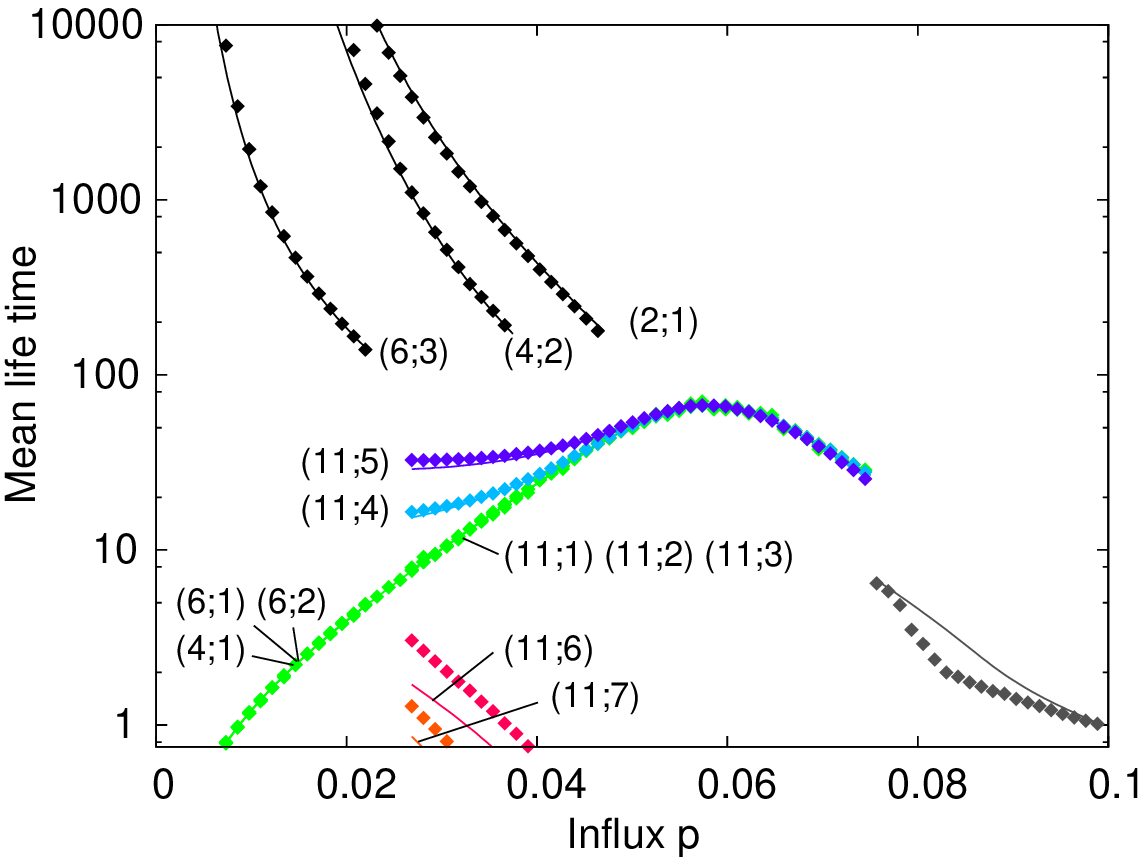}
\includegraphics[width=1.0\columnwidth]{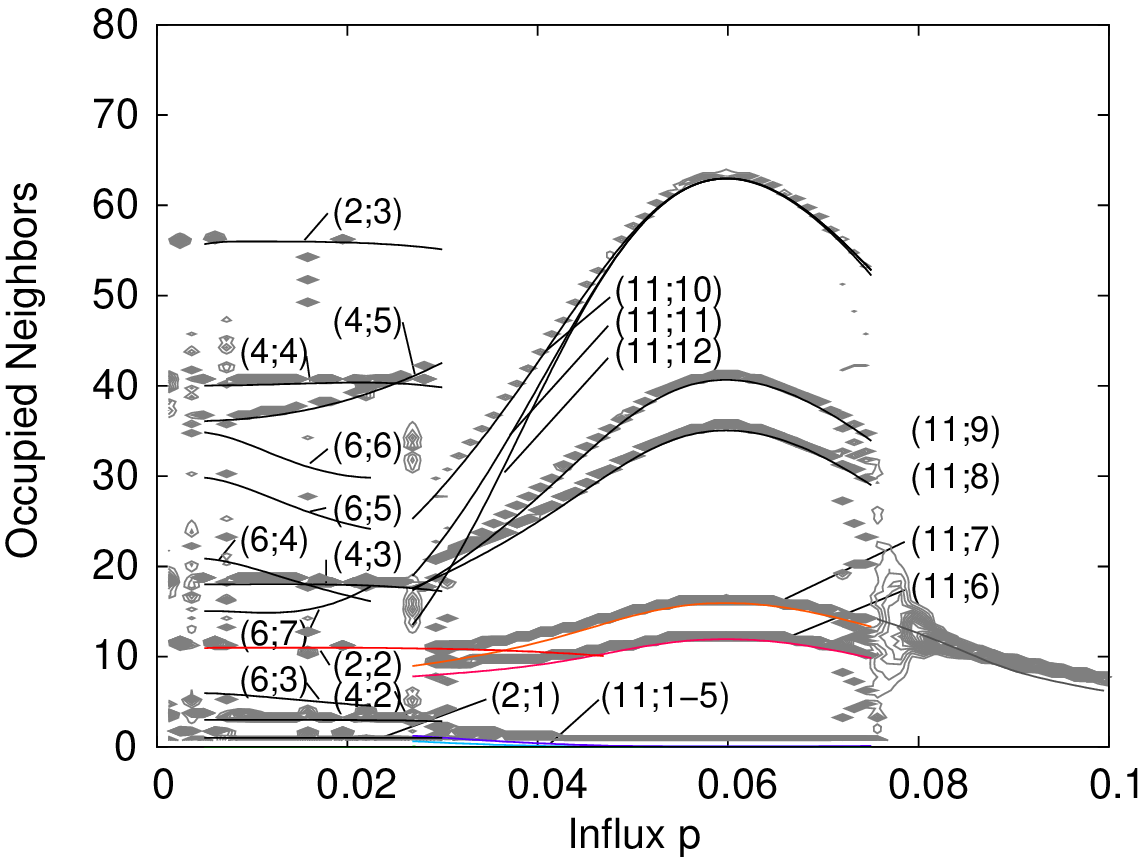}
\includegraphics[width=1.0\columnwidth]{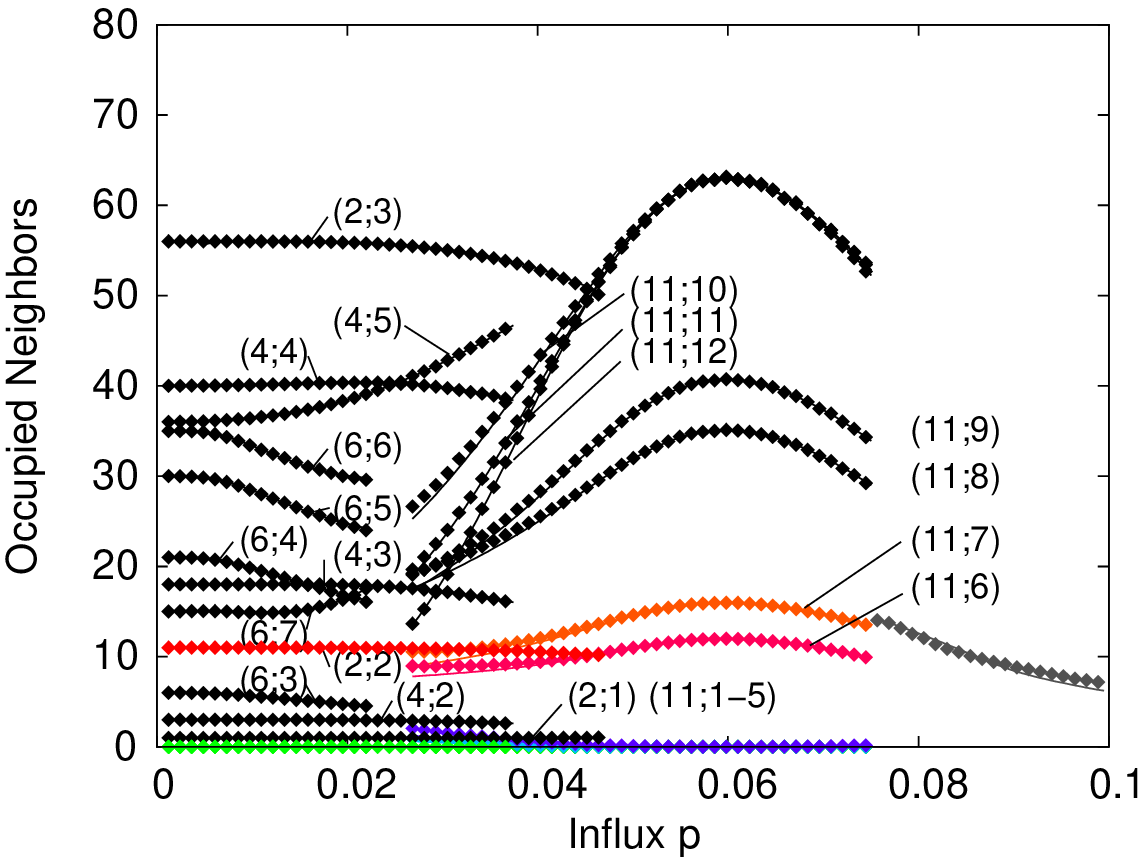}
\caption{(Color online) Statistical characteristics of nodes (mean occupation, mean life time, and mean number of occupied neighbors) vs influx probability $p$, in steps of $\Delta p=5/2^{12}$, obtained in mean field theory ({\em solid lines}) and by simulation.  {\em Left}: Simulations with 500,000 iterations starting from the empty base graph.  The {\em thin gray lines} are isolines (with increment 40) of histograms counting the frequency of nodes with a given characteristics after reaching the steady state, cf. Fig.~4 in \cite{STB}.  {\em Right}: Simulations with 200,000 iterations started from a snapshot of the respective pattern. The {\em symbols} represent averages over nodes which are members of the same  group. \\
The mean field results for nodes of group $S_g$ in a pattern module of dimension $d_M$ are given by solid lines labelled by $(d_M;g)$. They follow the ridges of the histograms ({\em left}) and coincide in most cases with the simulation of the prepared patterns ({\em right}). Groups with mean occupation near zero are not labelled. For further discussion see text.}
\label{fig:mfavssim} 
\end{figure*}

In Fig.~\ref{fig:mfavssim} we compare mean field results with the
simulation. In the left column we plotted isolines of histograms of
$\overline{n}(v)$, $\overline{n}(\partial v)$ and $\overline{\tau}(v)$ as a function
of $p$, the simulations were started from empty base graphs. In the
right column we used snapshots of the different patterns as initial
conditions. 

In the simulations started from the scratch for different values of
the influx $p$, the system evolved randomly and reached a steady state
with a stable architecture. In the steady state we measured
$\overline{n}(v)$, $\overline{n}(\partial v)$ and $\overline{\tau}(v)$ for all nodes
during an observation time of $500,000$ iterations.  For each choice
of $p$ we created histograms of the frequency of nodes against
one of the three statistical node characteristics. In the figure we show the isolines of a compilation of all
histograms in a range of $p \in [0,0.1]$. The observed aggregations of
nodes indicate the self-organization of the nodes into persisting
groups as well as the statistical differences of the groups in
dependence of $p$.

In general, the mean field results show a good agreement with the simulation. For $d_M=2$ we took the correlation correction into account, as described in Sec.~\ref{sec:correlation}.

Mean field results for some patterns are not matched by a
corresponding peak in the histograms. These patterns are seldom
reached in simulations starting from the scratch, because their basin
of attraction is too far away from the initial empty configuration or
behind a high barrier which is difficult to be passed.

The peaks which are not matched by a mean field result may occur if a
pattern does not persist over the observation time. This affects the
time average of the node statistics, see, for example, the left column at
$p\approx 0.0275$.

In the plots we also see that the character of the patterns changes
with increasing $p$. For moderate influx, $p \lessapprox 0.03$, we
identified almost static patterns associated with modules
$d_M=2,4,6$. 

For $p \approx 0.03 \dots 0.08$ there are no permanently
occupied nodes, the pattern is dynamic, but a well defined group
structure with $d_M=11$ exists, cf. Fig.~\ref{fig:scheme12g}. This is the most interesting architecture as explained above, cf. also \cite{BB03,STB}. The basin of attraction of the corresponding fixed point of Eq.~\ref{eq:scc} is large compared to the static patterns.

For $p \gtrapprox 0.075$ perturbations due to the random influx become
so large, that the orientation of the pattern module in the base graph
starts changing. Mathematically, these reorientations are
rotations \cite{Werner10}. After such a rotation there still exist 12 groups, but
their identification by temporal averaging of node characteristics is
impossible. The mean field theory does not consider rotations. It rather assumes
that the nodes always remain in their groups. However, we can compare averages over the whole graph from the simulations with the mean field results averaged over all groups
\begin{equation}
  \langle x \rangle_G = \frac{1}{|G|} \sum_{g=1}^{d_M+1} |S_g| x_g \,,
\end{equation}
where $x_g$ denotes one of the quantities $n_g, \tau_g,$ or $n(\dvg)$. These averages are in good agreement, see plots of mean occupation and occupied neighbors in Fig.~\ref{fig:mfavssim}.

We believe that these rotations of the pattern are the origin of the
difference between mean field results and simulation in the mean life
time plot at $p\approx 0.85$. Due to a rotation many
previously occupied nodes become stable holes. Thus, significantly
many nodes do not reach their usual life expectation.

The fact that we do not observe all patterns in the simulations
starting from an empty base graph, is a motivation to prepare the
initial conditions. In the right column of Fig.~\ref{fig:mfavssim} we compare the mean
field results to node statistics from simulations with different
predetermined initial architecture. The observation time was 200,000
iterations.
The picture proves that for the same value of $p$ different patterns
can persist, however, with different stability. The plotted lines of
a pattern end at the value of $p$ for which the pattern becomes
unstable.


\section{Conclusions\label{sec:conclusions}}

We considered a minimalistic model \cite{BB03} to describe the random evolution of
the idiotypic network which is, given very few model parameters,
mainly controlled by the random influx of new idiotypes and the
disappearance of not sufficiently stimulated idiotypes.  Numerical
simulations have shown that after a transient period a steady state is
achieved. Depending on the influx and on other parameters, the
emerging architecture can be very complex. Typically, groups of nodes
can be distinguished with clearly distinct statistical properties.
These groups are linked together in a characteristic way.

We achieved a detailed analytical understanding of the building
principles of these very complex structures emerging during the random
evolution \cite{STB}. Modules of remarkable regularity serve as building blocks
of the complex pattern. We can calculate for instance size and
connectivity of the idiotype groups in perfect agreement with the
empirical findings based on numerical simulations.

In this paper we developed a modular mean field theory which allows to compute the statistical characteristics of a variety of patterns given their architecture. The results are in very good agreement with the simulations. We thus have both structural and statistical information.
The quantities usually considered in the context of networks, like
degree distribution, cluster coefficient, centrality, etc., are
coarser than those investigated here and they do not reveal the
details of the architecture.

For certain groups of nodes, namely singletons and the self-stabilizing core in static patterns, which decouple in good approximation from the rest, the mean field treatment can be simplified considerably. For the pattern consisting of pairs of occupied neighbors, the 2-cluster pattern, the naive mean field theory fails. Therefore, we extended the mean field theory to include correlations. Also for the $d_M=11$ architecture correlations can be taken into account, which requires an intricate analysis of the building principles but is rewarded with an almost perfect match between theory and simulations \cite{KB}.

The modular mean field theory can be applied to calculate statistical properties of other stationary networks with known modular architecture.

\appendix*

\section{Stability\label{sec:stability}}

Stability of patterns is a central and complex question. Here we restrict ourselves to discuss the stability of the occupation state of a single node depending on the state of its neighborhood and the influx. We quantify this stability in terms of the probability that a node changes its occupation, i.e. to occupy empty nodes or to clear occupied nodes, respectively.

In the following we calculate the probability to change the occupation of a node $v$ given the number of occupied neighbors $n(\partial v)$. we distinguish three cases.\\
(1) $n(v)\!=\!0$, $n(\partial v)\! \leqslant\! t_U$. This is an unstable hole, because it can be occupied by the influx and will survive if $n(\partial \tildev) \in [t_L,t_U]$ after the influx. The probability of occupation and survival is
\begin{equation}
  P_1 (n(\partial v)) = p \sum _{k=t_L-n(\partial v)}^{t_U-n(\partial
    v)} P_{\kappa-n(\partial v), p} (k) \,, \label{eq:stability1}
\end{equation}
where $p$ is the probability that $v$ itself becomes occupied, and the
sum gives the probability that the number of occupied neighbors lies
in the window, i.e. $k \in [t_L-n(\partial v), t_U-n(\partial v)]$
empty neighbors become occupied. $P_{N, p}(k) = {N
  \choose k} p^k (1-p)^{N-k}$ is the binomial
distribution, where $P_{N, p}(k) \equiv 0$ if $k<0$.\\
(2) $n(v)\!=\!1$, $n(\partial v)\! \leqslant \! t_U$. This is an occupied node, including the case of an isolated node. The deletion of an occupied node requires that after the influx $\tilde{n}(\partial v)$ is outside the window. For $n(\partial v) < t_L$ the occupation of $k \in [0, t_L-n(\partial v)-1]$ or $k \in [t_U - n(\partial v) +1,
\kappa-n(\partial v)]$ empty neighbors is required. For $n(\partial v)
\in [t_L,t_U]$ the occupation of $k \in [t_U - n(\partial v) +1,
\kappa-n(\partial v)]$ is needed. This yields
\begin{equation}
  P_2 (n(\partial v)) = \left( \sum _{k=0}^{t_L-n(\partial v)-1} \!\!+\!\!
    \sum _{k=t_U-n(\partial v)+1}^{\kappa-n(\partial v)} \right)
  P_{\kappa-n(\partial v), p} (k) \,, \label{eq:stability2}
\end{equation}
where $\sum _{k=a}^{b} \dots \equiv 0$ if $b<a$.

The changes of occupation in cases (i) and (ii) are possible within one iteration. There are more complex scenarios which require more iterations. All occupation changes which demand the deletion of neighbors need at least two iterations, for instance the occupation of a stable hole. Such occupation changes also involve the neighbors of the neighbors. As an additional sophistication, we have to regard, that two neighbors of $v$ have overlapping neighborhoods, they have at least the node $v$ in common.\\
(3) $n(v)\!=\!0$, $n(\partial v)\! >\! t_U$. This is a stable hole. During the first
iteration the influx may increase the number of occupied neighbors by
$k \in [0, \kappa-n(\partial v)]$, such that the number of occupied
neighbors becomes $\tilde{n}(\partial v) = n(\partial v)+k$. The
probability for this process is $P_{\kappa-n(\partial v), p}(k)$. In
order to achieve $n'(\partial v) \leqslant t_U$, $l$ nodes of
$\partial v$ must be emptied, $l \in [\tilde{n}(\partial v)-t_U,
\tilde{n}(\partial v)]$.  The probability to empty one occupied node
$w \in \partial v$ is given by $P_2(n(\partial w))$, cf. Eq.
(\ref{eq:stability2}). However, the probability to empty $l$ occupied
nodes in $\partial v$ is not simply $\prod_{i=1}^l P_2(n(\partial
w_i))$, because the neighborhoods $\partial w_i$ are not disjoint.
Neglecting this correlation, one can proceed with the second iteration
as follows. We have after the first iteration $n'(v)=0$ and $n'(\partial v) =
n(\partial v) + k - l \leqslant t_U$. Thus, the second iteration is
simply case (1), i.e. the occupation of an unstable hole.  Collecting the
probabilities of the consecutive steps for all possible choices of $k$
and $l$ gives then
\begin{align} \label{eq:stability3} P_3(\partial (\partial v))=&\sum
  _{k=0} ^ {\kappa-n(\partial v)}
  P_{\kappa-n(\partial v), p} (k)\nonumber \\
  & \times \sum _{l=n(\partial v)+k-t_U}^{n(\partial v)+k} \sum
  _{\{C_l\}} \prod
  _{w\in C_l} P_2 (n(\partial w)) \nonumber \\
  & \times P_1(n(\partial v)+k-l)\,,
\end{align}
where $\sum_{\{C_l\}} \dots$ denotes the sum over all choices of $l$
occupied nodes of $\partial v$ which are emptied.

For our parameter setting, and generally if $t_L \ll t_U \ll \kappa$, we find $P_1 \geqslant P_2 \geqslant P_3$. Stable holes are the nodes with the most stable occupation state.  We expect that a pattern is stable, i.e. its nodes preserve their statistical properties for a long time, if it has sufficiently many stable holes.

\begin{acknowledgments}
\hyphenation{IMPRS}
Thanks is due to Andreas K\"uhn, Heinz Sachsenweger, Mario Th\"une, Benjamin Werner, and Sven Willner for valuable comments. H.S. thanks the IMPRS Mathematics in the Sciences and the Evangelisches Studienwerk Villigst e.V. for funding.
\end{acknowledgments}


\end{document}